\documentclass[journal]{IEEEtran}
\usepackage{cite}
\usepackage{amsmath,amssymb,amsfonts}
\usepackage{graphicx}
\usepackage{textcomp}
\usepackage{xcolor}
\usepackage{comment}
\usepackage{algorithm}
\usepackage{algpseudocode}
\usepackage{babel}
\usepackage{multirow}

\newcommand{\h}{\boldsymbol{h}_k}

\begin{document}

\title{Detection for Flat-Fading Channels based on Expectation Propagation\\}

\author{Elisa~Conti,~Amina~Piemontese,~Giulio~Colavolpe, and~Armando~Vannucci%

\thanks{The authors are with the Department of Engineering and Architecture, University of Parma, 43124 Parma, Italy (e-mails: \{elisa.conti, amina.piemontese, giulio.colavolpe, armando.vannucci\}@unipr.it).}
}

\maketitle

\begin{abstract}
This paper aims at tackling the problem of signal detection in flat-fading channels. In this context, receivers based on the \textit{expectation propagation} framework appear to be very promising although presenting some critical issues. We develop a new algorithm based on this framework where, unlike previous works, convergence is achieved after
a single forward-backward pass, without additional inner detector iterations. 
The proposed message scheduling, together with novel
adjustments of the approximating distributions' parameters, allows to obtain significant performance advantages with respect to the state-of-the-art solution. Simulation results show the applicability of this algorithm when sparser pilot configurations have to be adopted and a considerable gain compared to the current available strategies. 
\end{abstract}


\section{Introduction}

\IEEEPARstart{D}{etectors} for flat Rayleigh fading channels have been widely studied in the literature (e.g., see~\cite{NiShRiLi05, QiMi07, BaPiCo09} and references therein).
This kind of channels is commonly adopted because it properly represents real scenarios, such as narrowband communications over multipath wireless channels. The framework based on factor graphs (FGs) and the sum-product algorithm (SPA) provides a powerful tool for the accomplishment of both detection and decoding~\cite{KsFrLo01}. 
However, when this framework is applied jointly to detection and channel estimation, discrete and continuous random variables appear in the FG and the SPA becomes impractical. 

A common approach foresees the use of canonical distributions~\cite{KsFrLo01}. Focusing on the FG methodology and on the choice of a predefined family which allows some kind of parametrization, different approaches have been analyzed~\mbox{\cite{BaPiCo09,NiShRiLi05}}. A canonical distribution that is commonly selected for the messages in this scenario is the Gaussian one. 
Due to the presence of both discrete and continuous random variables, the \textit{observation} message is typically represented by a multimodal distribution. One of the main differences among the various methods lies in the technique followed for its approximation.
In~\cite{NiShRiLi05}, the authors proposed to set the mean to the empirical average of the Gaussian mixture and fix the variance to that of the additive white Gaussian noise (AWGN).
In~\cite{BaPiCo09}, the approximating probability density function (pdf) is the Gaussian that minimizes the Kullback-Leibler (KL) divergence from the true message~\cite{CoTh91} and whose variance depends on the fading sample (i.e., on a variable node), which complicates the resulting algorithm. Such a dependence is removed by substituting the fading sample with its estimate performed at the previous iteration. 
In the separate detection and decoding scenario, this approach coincides with the one of~\cite{NiShRiLi05}. 
By reducing the multimodal distributions to single Gaussian ones, the forward and backward recursions of the algorithm coincide with those of a Kalman smoother~\cite{JoBi01}. Although this approach, based on message projection, is effective in pilot-aided transmission, it eventually fails when the spacing between pilot blocks is longer than a maximum accepted distance, as we will demonstrate in the numerical results.

In order to overcome this limitation, in this paper we focus on a different kind of receiver, which is based on the expectation propagation (EP) framework proposed in~\cite{Mi01} and generalized in~\cite{Mi05}. This message-passing algorithm is based on approximating the distributions of the variables of interest, rather than those of the exchanged messages. This way, a \textit{prior belief} coming from the rest of the FG drives the approximation of intractable messages.
The combination of the SPA rules with the approximation deriving from EP leads to a hybrid framework which, in the literature, is commonly referred to as EP-belief propagation (BP). This framework has been deeply studied and analyzed in different scenarios, including, e.g., the transmission over phase noise channels~\cite{CoCoPiVa23}, the MIMO sparse codes multiple access detection~\cite{YuWuGu18} or joint channel estimation and decoding for faster-than-Nyquist signaling~\cite{WuYuGu17} over frequency selective fading channels.
As it is inherent to the EP update strategy, improper distributions, e.g., Gaussians with a negative variance, could appear. 
While in~\cite{QiMi07} the presence of improper distributions is accepted, and no additional operations are introduced, different approaches are followed by other authors, consisting in either rejecting the message updates that produce improper distributions~\cite{Se03} or, in the case of a multivariate normal distribution, operating on the covariance matrix through a replacement of negative eigenvalues~\cite{GeVe14}. 

In \cite{QiMi07}, the EP framework has been applied to a flat-fading channel, modeled as a complex autoregressive moving-average process. The authors proposed an EP smoothing approach where inner detector iterations are required to achieve convergence and it is tested using differential encoding. 

However, when using a more powerful encoder, hence working in a lower signal-to-noise ratio (SNR) regime, the approach in \cite{QiMi07} is not equally effective and this is the focus of our investigation. 

The main contributions of this letter can be summarized as follows.
\begin{itemize}
    \item We propose an implementation of EP-BP that overcomes
    its limitations and reduces the algorithmic complexity.
    \item An improved scheduling which allows to achieve convergence without the aid of multiple inner detector iterations (\textit{iterative refinement}~\cite{Mi01}) is adopted.
    \item A new technique for balancing the variance of the involved Gaussian distributions is introduced.
    \item The problem of improper distributions is addressed, the scenarios where they can compromise the algorithm operations are identified and a novel strategy to handle these events is presented.
\end{itemize}
Finally, we demonstrate through simulations that the proposed techniques present a superior performance, in terms of bit-error rate (BER), with respect to state-of-the-art EP implementations~\cite{QiMi07}. Moreover, the proposed solution ensures a higher robustness against concentrated pilot distributions, laying the foundation for overcoming the limitations related to the classical and well-established Kalman filter.

The rest of the paper is organized as follows. Section~\ref{sec:model} introduces the system model. Section~\ref{sec:factorization_and_EP} is devoted to the description of the EP algorithm. The proposed techniques are introduced in Section~\ref{sec:prop_alg}. Numerical results are presented in Section~\ref{sec:results} while Section~\ref{sec:conclusions} collects some concluding remarks.

\section{System Model}\label{sec:model}
We consider the transmission of a sequence of $K$ complex coded symbols $\boldsymbol{c}=\left(c_0,c_1,\dots,c_{K-1}\right)$ over a flat fading channel. 
The sequence $\boldsymbol{c}$ is the result of the 
binary encoding, through the code $\mathcal{C}$ and 
modulation over an \mbox{$M$-ary} constellation ($\mathcal{X} = \left\{x_1, \dots, x_M\right\}\subset \mathbb{C}$) of the information bit sequence $\boldsymbol{b}$. 
Assuming linear modulation with Nyquist shaping pulses, matched filtering, and channel variations slow enough so that no intersymbol interference arises, the discrete-time received signal can be expressed as
\begin{equation}
r_k = g_kc_k + n_k,\quad k = 0,1,\dots, K-1
\end{equation}
where ${n_k}$ is the AWGN and ${g_k}$ is the fading process. The vector of noise samples, $\boldsymbol{n}=\left(n_0,n_1,\dots,n_{K-1}\right)$,\footnote{The following notation is used throughout the paper. Bold lower-case letters ($\boldsymbol{a}$) denote vectors, bold upper-case letters ($\boldsymbol{A}$) denote matrices. The superscripts $(\cdot)^T$ and $(\cdot)^H$ denote the transpose and the Hermitian of a matrix or vector, respectively. Given a square matrix $\boldsymbol{B}$ with dimension $N_B\times N_B$, the generic $i,j$ element is denoted by $\boldsymbol{B}(i,j)$, $i,j=1,\dots,N_B$.} has independent and identically distributed (i.i.d) complex circularly symmetric components with $n_k\sim\mathcal{N}_\mathbb{C}\left(0,N_0\right)$.
\footnote{A complex circularly symmetric Gaussian random variable $v$ with mean $\mu$ and variance $2\sigma^2$ is denoted by $v\sim\mathcal{N}_\mathbb{C}\left(\mu, 2\sigma^2 \right)$. The corresponding probability density function (pdf), with argument $x$, is denoted as $g_\mathbb{C}\left(x,\mu, 2\sigma^2 \right)$. We adopt a similar notation for a Gaussian vector, whose pdf is written as $g_\mathbb{C}\left(\boldsymbol{x},\boldsymbol{\mu}, \boldsymbol{C} \right)$, where the inverse of the covariance matrix, $\boldsymbol{\Sigma}=\boldsymbol{C}^{-1}$, is called \textit{precision}.}

The fading channel follows Clarke's isotropic scattering model and the fading coefficients $\boldsymbol{g}=\left(g_0,g_1,\dots,g_{K-1}\right)$ form a sequence of zero-mean complex Gaussian random variables with unit variance, statistically independent of both $\boldsymbol{c}$ and $\boldsymbol{n}$, with autocorrelation sequence 
\begin{equation}\label{eq:Clarke}
    R_g(m)=E\left\{{g_{k+m}g^{*}_k}\right\}=J_0(2\pi f_DTm)
\end{equation}
where $J_0(\cdot)$ is the zero-th order Bessel function, $T$ is the symbol period, and $f_D$ is the Doppler spread.
At the receiver, the adopted model for the fading is an autoregressive process of order $N$, AR($N$),
so that $\boldsymbol{g}$ is approximated by the process $\boldsymbol{h}=\left(h_0,h_1,\dots,h_{K-1}\right)$, which obeys\footnote{We assume that $h_l = 0$ for $l<0$.}
\begin{equation}\label{eq:ARp_model}
h_k = \sum_{n = 1}^{N} \rho_n h_{k-n} + \nu_k = \boldsymbol{\rho}\boldsymbol{h}^T_{k-1} + \nu_k\, ,
\end{equation}
where $\nu_k \sim\mathcal{N}_\mathbb{C}\left(0,2\sigma^2_\nu\right)$, $\boldsymbol{h}_{k-1}=\left(h_{k-1},\dots,h_{k-N}\right)$ and \mbox{$\boldsymbol{\rho}=\left(\rho_1,\rho_2, \dots,\rho_N\right)$} are the real AR($N$) coefficients. 
Clearly, there is a channel mismatch between the AR($N$) process $\left\{h_k\right\}$ and the one modelled by (\ref{eq:Clarke}), hence it is critical to properly set the values of the coefficients $\boldsymbol{\rho}$ in (\ref{eq:ARp_model}) and of the random increment's variance $2\sigma^2_\nu$. 
A classical approach in the literature~\cite{NiShRiLi05,HeCh00} consists in obtaining the coefficients $\boldsymbol{\rho}$ by solving the Yule-Walker (YW) equations \mbox{$R_h(m)=R_g(m)$}, for $m = 1,2,\dots, N$, and computing $\sigma^2_\nu$ as a function of $\boldsymbol{\rho}$ through the constraint $R_h(0)=R_g(0)$. 
However, in the present context, it turns out that this solution is not the best choice to approximate the actual fading statistics~\cite{BaPiCo09}. Therefore, we will select the value of $\sigma^2_\nu$ that optimizes the performance, as detailed in Section~\ref{sec:results}.

\section{The Expectation Propagation Framework}\label{sec:factorization_and_EP}

Solving the fading estimation problem through message-passing algorithms requires the factorization of the joint posterior pdf of the information bits and the fading coefficients, given the received samples. From the system model described above, it can be expressed as
\begin{equation}\label{eq:joint_pdf}
\begin{split}
    & P(\boldsymbol{b}, \boldsymbol{c}, \boldsymbol{h}|\boldsymbol{r}) \propto p(\boldsymbol{r}|\boldsymbol{c},\boldsymbol{h}) p(\boldsymbol{h})P(\boldsymbol{c}|\boldsymbol{b})P(\boldsymbol{b}) \\
    & \!\propto \!I[\boldsymbol{c}=\mu_\mathcal{C}(\boldsymbol{b})] p(h_0)\!\!\prod_{k=1}^{K-1}\!\!p_\nu(h_k\!-\boldsymbol{\rho}\boldsymbol{h}^T_{k-1})\!\! \prod_{k=0}^{K-1}f_k(c_k,h_k) ,
\end{split}
\end{equation}
where the information bits are uniformly and identically distributed, $\mu_\mathcal{C}$ is the  function that maps binary information messages $\boldsymbol{b}$ into the codewords $\boldsymbol{c}$, and $ I[ \cdot ]$ is an indicator function equal to $1$ when $\boldsymbol{c}=\mu_\mathcal{C}(\boldsymbol{b})$ and to zero otherwise. 
The factorizations of the pdfs $p(\boldsymbol{h})$ and $p(\boldsymbol{r}|\boldsymbol{c},\boldsymbol{h})$ in~(\ref{eq:joint_pdf}) come, respectively, from the assumption that the fading process follows the AR($N$) model~(\ref{eq:ARp_model}) and from the fact that the AWGN channel is memoryless, given $\boldsymbol{h}$. 

\begin{figure}
    \centering
    \includegraphics[width=0.8\columnwidth]{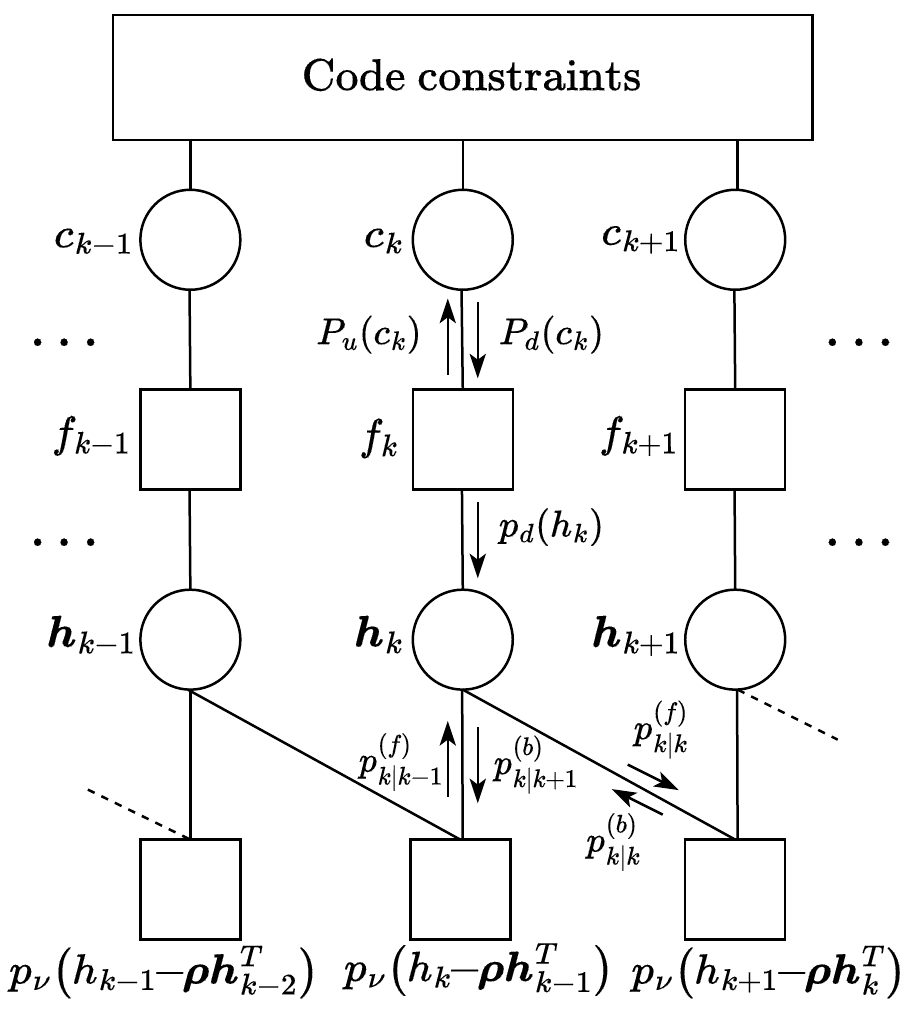}
    \caption{FG corresponding to~($\ref{eq:joint_pdf}$).}
    \label{fig:FG}
\end{figure}

In (\ref{eq:joint_pdf}), we defined 
\begin{equation}
   p_\nu(h_k-\boldsymbol{\rho}\boldsymbol{h}^T_{k-1})  \triangleq p(h_k|\boldsymbol{h}_{k-1}) = g_\mathbb{C}(h_k,\boldsymbol{\rho}\boldsymbol{h}^T_{k-1},2\sigma^2_\nu) \\
\end{equation}
\begin{equation}
   \hspace{-0.11cm}f_k(c_k,h_k)  \triangleq p(r_k|c_k,h_k) = g_\mathbb{C}(r_k,h_{k}c_k,N_0)
\end{equation}
as the functions associated with the factor nodes in the FG representation of the joint pdf ($\ref{eq:joint_pdf}$), which is reported in Fig.~\ref{fig:FG}.

Due to the presence of continuous random variables in the FG, we follow the commonly adopted technique based on canonical distributions~\cite{KsFrLo01}. In this way, only the parameters characterizing a specific pdf within the family are exchanged in the FG. From~(\ref{eq:ARp_model}), the distributions that better approximate those of the fading samples are the Gaussian ones. According to the SPA rules~\cite{KsFrLo01}, which we assume the reader to be familiar with, the \textit{observation} message, i.e., the message from factor node $f_k$ to variable node $\boldsymbol{h}_k$ is 
\begin{equation}\label{eq:pd}
     p_d(h_k) \propto \sum^{M}_{m = 1}P_d(c_k = x_m)f_k(c_k=x_m,h_k) \, , 
\end{equation}
hence it is the mixture of $M$ Gaussian distributions and its exact propagation would be infeasible.  
Depending on how the message in~(\ref{eq:pd}) is treated, different algorithms can be derived.

A straightforward projection of the mixture message~(\ref{eq:pd}) onto a single Gaussian yields the Kalman filtering approach of~\cite{NiShRiLi05,BaPiCo09}. 
Once the projection is performed, all the subsequent messages that are sent along the Markov chain at the bottom of Fig.~\ref{fig:FG} belong to the same approximating family because of the closure of the Gaussian family, under both the multiplication and the convolution operations~\cite{NiShRiLi05,KsFrLo01}. 

A different possibility consists in approximating the distributions of variable nodes, rather than the messages, and this is the key idea on which the EP framework is based.
The approximation of $p_d(h_k)$ in (\ref{eq:pd}) is thus performed in EP, under the influence of a \textit{temporary prior} distribution which is the product of the incoming messages to the variable node $\boldsymbol{h}_k$ from the lower part of the FG. 
When, on the contrary, a message projection is carried out as in the Kalman filtering approach, so that no prior information is exploited, this is sometimes called \textit{transparent} propagation algorithm \cite{VannucciColavolpePecoriVeltri_ISPLC_2019, VannucciColavolpeVeltri_CommLett_2020_ImpNoise}.

EP is a message-passing algorithm whose approximations are generated by projecting  the variable distribution on a fully-factorized exponential family $\mathcal{F}$, where the projection is performed by minimizing the Kullback-Leibler (KL) divergence. 
These two features allow to compute the projection by solving a set of \textit{moment matching} equations~\cite{Mi05}. 
In the present case of a Gaussian approximating family, the distribution resulting from the projection is thus the unique Gaussian whose mean, variance and scale factor equal those of the distribution to be approximated.

The specific algorithmic steps of EP are described in the following. 
Starting from (\ref{eq:pd}), the message $p_d(h_k)$ can be rewritten as a function of the channel parameter $h_k$ as follows 
\begin{equation}\label{eq:pd_hk}
    p_d(h_k) =\sum_{m=1}^{M}P_m\,g_\mathbb{C}(h_k,\eta_m,2\sigma^2_m)\,,
\end{equation}
where
\begin{equation}\label{eq:P_m}
    P_m =\frac{P_d(c_k = x_m)}{\left|x_m\right|^2}\hspace{2mm},\hspace{2mm}
    \eta_m =\frac{r_k}{x_m}\hspace{2mm},\hspace{2mm}
    2\sigma^2_m =\frac{N_0}{\left|x_m\right|^2}\hspace{2mm}.
\end{equation}
The dependence of $P_m$ and $\eta_m$ on the time epoch $k$ is omitted for simplicity of notation.
The temporary prior distribution $p^{(p)}(\h)$ is  defined as
\begin{equation}\label{eq:prior_distribution}
\begin{split}
    &p^{(p)}(\boldsymbol{h}_k) \triangleq p^{(f)}_{k|k-1}(\boldsymbol{h}_k)p^{(b)}_{k|k+1}(\boldsymbol{h}_k)\\
                &=g_\mathbb{C}\left(\boldsymbol{h}_k,\boldsymbol{\eta}_{k|k-1},\boldsymbol{C}_{k|k-1}\right)g_\mathbb{C}\left(\boldsymbol{h}_k,\boldsymbol{\eta}_{k|k+1},\boldsymbol{C}_{k|k+1}\right) \\
                 &\propto g_\mathbb{C}\left(\boldsymbol{h}_k,\boldsymbol{\eta}_{p,k},\boldsymbol{C}_{p,k}\right)\,,
\end{split}
\end{equation}
where the $\propto$ sign derives from the messages normalization, and, by defining $\boldsymbol{\Sigma}_{f,k}\triangleq \boldsymbol{C}_{k|k-1}^{-1}$ and $\boldsymbol{\Sigma}_{b,k}\triangleq \boldsymbol{C}_{k|k+1}^{-1}$ as the precisions of the forward and backward Gaussian messages, respectively, and $\boldsymbol{\Sigma}_{p,k}\triangleq \boldsymbol{C}_{p,k}^{-1}$, the following Gaussian product rules can be applied
\begin{align}\label{eq:prior precision}    \boldsymbol{\Sigma}_{p,k}&=\boldsymbol{\Sigma}_{b,k}+\boldsymbol{\Sigma}_{f,k} \,,\\\label{eq:prior mean}  
{\boldsymbol{\Sigma}_{p,k}} \boldsymbol{\eta}_{p,k}&= \boldsymbol{\Sigma}_{f,k}\boldsymbol{\eta}_{k|k-1}+\boldsymbol{\Sigma}_{b,k}\boldsymbol{\eta}_{k|k+1} \, .
\end{align}
The marginal distribution of $\boldsymbol{h}_k$ is given by the product of all the incoming messages to the corresponding variable node, according to the SPA rules~\cite{KsFrLo01}
\begin{equation}\label{eq:P_hat_final}
\begin{split}
    \hat{p}(\boldsymbol{h}_k) &= p_d(\boldsymbol{h}_k)p^{(f)}_{k|k-1}(\boldsymbol{h}_k)p^{(b)}_{k|k+1}(\boldsymbol{h}_k) \\
    &=\sum_{m=1}^{M}P_m\,g_\mathbb{C}\left(h_k,\eta_m,2\sigma^2_m\right)g_\mathbb{C}\left(\boldsymbol{h}_k,\boldsymbol{\eta}_{p,k},\boldsymbol{C}_{p,k}\right) \\
    &=\sum_{m=1}^{M}P_m w_m\,g_\mathbb{C}\left(\boldsymbol{h}_k,\boldsymbol{\eta}_{MIX,m},\boldsymbol{C}_{MIX,m}\right) \,,
\end{split}
\end{equation}
where $w_m$ in~(\ref{eq:P_hat_final}) is the symbol-dependent weight that multiplies the Gaussian distribution corresponding to the $m$-th symbol in~(\ref{eq:pd}) and is defined as
\begin{equation}\label{eq:w_m}
    w_m=g_\mathbb{C}\left(\eta_m,\boldsymbol{a}^T\boldsymbol{\eta}_{p,k},2\sigma^2_m+\boldsymbol{a}^T\boldsymbol{C}_{p,k}\boldsymbol{a}\right)\,,
\end{equation}
where $\boldsymbol{a}=\left(1,0,\dots,0\right)^T$ is a column vector of dimension $N$ while $\boldsymbol{C}_{MIX,m}$ and $\boldsymbol{\eta}_{MIX,m}$, after some calculations, here omitted for brevity, result:
\begin{align}\label{eq:eta_MIX}
   &\boldsymbol{\eta}_{MIX,m} = \boldsymbol{\eta}_{p,k} + \boldsymbol{J}_{m,k}\left(\eta_m-\boldsymbol{a}^T\boldsymbol{\eta_{p,k}}\right)\,,\\
   \label{eq:C_MIX}
  &\boldsymbol{C}_{MIX,m} =\boldsymbol{C}_{p,k}-\boldsymbol{J}_{m,k}\,\boldsymbol{a}^T \boldsymbol{C}_{p,k}\,,
\\ \label{eq:J}
    &\boldsymbol{J}_{m,k} = \boldsymbol{C}_{p,k}\,\boldsymbol{a}\left(2\sigma^2_m+\boldsymbol{a}^T\boldsymbol{C}_{p,k}\,\boldsymbol{a}\right)^{-1}\,.
\end{align}

Notice that the variable distribution in~(\ref{eq:P_hat_final}) is still a combination of $M$ Gaussians but the weight of each one, $w_m$, depends on the distance between the mean of that mode ($\eta_m$) and that of the temporary prior distribution. The closer they are, the more the corresponding mode is reinforced by the prior and vice versa, as per~(\ref{eq:w_m}).  

The approximation of the {mixture (\ref{eq:P_hat_final}) with a simple Gaussian $p^{(EP)}(\boldsymbol{h}_k) =\text{proj}_{KL}\left[\hat{p}(\boldsymbol{h}_k)\right]$ is performed, as usual in EP, by \textit{matching the moments} of the distributions. The zero-th order moment\footnote{We denote by $\mathbb{E}_{p}\left[x\right]$ the expectation of $x$ under the distribution $p(\cdot)$.} 
$\mathbb{E}_{p^{(EP)}}\left[ 1 \right]=\mathbb{E}_{\hat{p}}\left[ 1 \right]$ is simply the mass of the distribution, hence it would be necessary to set the mass of $p^{(EP)}(\boldsymbol{h}_k)$ to $Z=\sum_{m=1}^{M}P_m w_m$ that results from (\ref{eq:P_hat_final}). 
However, the mass of the messages is not significant in message-passing algorithms like SPA or EP, therefore we prefer to consider $p^{(EP)}(\boldsymbol{h}_k)$ with unit mass (hence a proper pdf) and to replace $\hat{p}(\boldsymbol{h}_k)$ in (\ref{eq:P_hat_final}) with its normalized version}
\begin{equation}
    \!\!\!p_h(\boldsymbol{h}_k) \!= \!\frac{\hat{p}(\boldsymbol{h}_k)}{Z} \!=\! \!\sum_{m=1}^{M}\!W_m g_\mathbb{C}\left(\boldsymbol{h}_k,\boldsymbol{\eta}_{MIX,m},\boldsymbol{C}_{MIX,m}\right)\,.
\end{equation}
where $W_m = Z^{-1}P_m w_m$ are defined as normalized weights.
The matching of the first order moment is thus performed by equating the mean of the two distributions
\mbox{$\mathbb{E}_{p^{(EP)}}\left[\boldsymbol{h}_k\right]=\mathbb{E}_{p_h}\left[\boldsymbol{h}_k\right]$}.
\begin{comment}
\begin{equation}
    i +\mathbb{E}_{p^{(EP)}}\left[h_k\right]=\mathbb{E}_{p_h}\left[h_k\right]\,.
\end{equation}
\end{comment}
It follows that the mean of $p^{(EP)}(\h)$ is
\begin{equation}\label{eq:eta_EP}
    \boldsymbol{\eta}_{EP} =\sum_{m=1}^{M}W_m\,\boldsymbol{\eta}_{MIX,m}\,.
\end{equation}
Then, by matching the second order moments, its covariance matrix is obtained as
\begin{equation}\label{eq:sigma_EP}
\boldsymbol{C}_{EP}\!=\!\!\sum_{m=1}^{M}\!W_m\!\left(\boldsymbol{C}_{MIX,m}+\boldsymbol{\eta}_{MIX,m}\boldsymbol{\eta}^H_{MIX,m}\right) - \boldsymbol{\eta}_{EP}\boldsymbol{\eta}^H_{EP}\,.
\end{equation}
\begin{comment}
Finally, the approximating distribution can be written as
\begin{equation}\label{eq:approx_distribution}
    p^{(EP)}(h_k)\propto g_\mathbb{C}\left(h_k,\eta_{EP},2\sigma^2_{EP}\right)\,.
\end{equation}
\end{comment}
Now, it is possible to compute the updated message 
\begin{equation}\label{eq:pd_approx}
    \Tilde{p}_d(\boldsymbol{h}_k)\propto g_\mathbb{C}\left(\boldsymbol{h}_k,\boldsymbol{\eta}_{d},\boldsymbol{C}_{d}\right)\,,
\end{equation}
which is transmitted from the factor node $f_k$ to the variable node $\boldsymbol{h}_k$ in Fig.~\ref{fig:FG}, dividing the approximating {Gaussian} $p^{(EP)}(\boldsymbol{h}_k)$ by the temporary prior distribution $p^{(p)}(\boldsymbol{h}_k)$ in~(\ref{eq:prior_distribution}). We obtain
\begin{align}\label{eq:Sigma_d}
   \boldsymbol{\Sigma}_d &= \boldsymbol{\Sigma}_{EP}-\boldsymbol{\Sigma}_{p,k}\,,\\\label{eq:eta_d}
    \boldsymbol{\Sigma}_d\boldsymbol{\eta}_d &= \boldsymbol{\Sigma}_{EP}\boldsymbol{\eta}_{EP}-\boldsymbol{\Sigma}_{p,k}\boldsymbol{\eta}_{p,k} \, .
\end{align}
The approximated message is then propagated following the SPA rules\,\cite{KsFrLo01}.
%----------------------------------------------------

\section{Proposed Algorithm}\label{sec:prop_alg}

In the considered scenario, the classical EP implementation~\cite{QiMi07} presents some critical issues which do not lead to a satisfactory performance. Therefore, in the following, we will present novel techniques which allow to overcome the limitations related to the standard EP structure.  

% ************************************************
\subsection{Damping and Boosting Factors}\label{subsec:damping}
In order to face the EP instabilities, as discussed in the literature, a number of solutions exists, including \textit{damping}~\cite{GeVe14,Se03}.
A dumped update of a pdf is a member of the approximating family characterized by a convex combination of new parameters (i.e., updated according to the current EP iteration) and old ones (i.e., from previous iteration).
However, unlike its common use, in the analyzed scenario the introduction of this correction factor is studied specifically to provide a performance improvement through the balancing of overly confident estimates without the involvement of old updates. 
In fact, in our implementation, the damping factor, $\xi_d$, which belongs to the interval $[0,1]$, is applied to the precision of the approximating message $\Tilde{p}_d({h}_k)$ only. In this way, the algorithm underestimates the updates reliability as commonly done in the literature when dealing with sub-optimal algorithms.
We propose
to set this factor to the maximum value of $P_d(c_k)$, i.e., the
probability mass function of the symbols. The main advantage of this approach is that a lower number of turbo iterations is required for achieving convergence.
At the first turbo iteration\footnote{We will refer to the iterations between detector and decoder as \textit{turbo} iterations.} (which is the only one in the separate detection and decoding case) this quantity is equal to $1/M$. On the contrary, from the $2$nd iteration ahead this value becomes a clearer and clearer representation of the decoder confidence about the transmitted symbols. It follows that the closer this value is to $1$, the higher the weight given to the approximating message after the KL projection.

Moreover, we propose an additional technique which allows to obtain a further performance improvement: the use of a \textit{boosting} factor for pilot symbols. 
Similarly to damping, the
precision of the distribution in~(\ref{eq:pd_approx}), whose expression is reported in~(\ref{eq:Sigma_d}), is multiplied by a factor that,
this time, is greater than $1$, so that its effect is opposite: it increases the confidence of the observation message in correspondence with a pilot symbol, 
i.e., when $p_d({h}_k)$ is a unimodal Gaussian distribution and, therefore, no approximation is performed. The selected value for the boosting factor is $\xi_b=2$.
% ************************************************

\subsection{Negative Variances}\label{subsec:neg_variances}

Due to the subtraction in (\ref{eq:Sigma_d}), it may occur that the algorithm generates negative variances, so that the approximation of the observation message, $\Tilde{p}_d({h}_k)$, is not necessarily a valid probability distribution~\cite{QiMi07}. This happens when the precision of the prior distribution is larger than the one of the marginal after the KL projection. Since the focus of EP is on the marginal distribution of the channel parameters, the algorithm could accept improper distributions as messages. However, when the scheduling described in Section~\ref{subsec:scheduling} is adopted for the separate detection and decoding case, the only message which can present such variances is the observation one.

When turbo iterations are considered, improper distributions could instead propagate giving rise to numerical problems. Hence, in this case, we propose an alternative approach to deal with negative variances, where the algorithm performs a message projection instead of a variable distribution projection. This leads to a significant performance improvement with respect to the propagation of improper distributions as shown in Section~\ref{sec:results}.
The proposed method consists in approximating the message $p_d(h_k)$ locally with a Gaussian distribution whose mean and variance are obtained through moment matching without the influence of the rest of the network.

The approximating message can be written as 
\begin{equation}
\!\hat{p}_d(h_k)=\text{proj}_{\text{KL}}\left[\sum_{m=1}^{M}\frac{P_d(c_m)}{\left|c_m\right|^2}\,g_\mathbb{C}\left(h_k,\frac{r_k}{c_m},\frac{2\sigma^2_n}{\left|c_m\right|^2}\right)\right]\,.
\end{equation}

With the constraint on the Gaussian approximating family, the new mean and variance will be
\begin{equation}\label{eq:pd_proj_mean}
   \hspace{-3.8cm}\eta_d = \sum_{m=1}^{M}\frac{P_d(c_k=x_m)}{\left|x_m\right|^2}\cdot\frac{r_k}{x_m}\,,\\
\end{equation}
\begin{equation}\label{eq:pd_proj_var}
    2\sigma^2_d = \sum_{m=1}^{M}\frac{P_d(c_k = x_m)}{\left|x_m\right|^2}\cdot\left(\frac{N_0}{\left|x_m\right|^2}+\left|\frac{r_k}{x_m}\right|^2\right)-\left|\eta_d\right|^2\,.
\end{equation}
Therefore, if, following the main procedure where the KL divergence minimization is performed on the marginal distribution, negative variances occur, the resulting observation message update is discarded and the new technique is applied.

\subsection{Scheduling}\label{subsec:scheduling}

We assume that the detector subgraph operates first by exchanging \textit{horizonal} messages along the Markov chain in the lower part of the FG shown in Fig.~\ref{fig:FG}. This message passing, implemented inside the detector, could be iterated with multiple forward-backward passes ($N_d$) before sending \textit{vertical} messages to the upper part of the FG where the decoder operates. An inner message passing procedure could take place also there until convergence or until a maximum number of decoder iterations ($N_c$) is reached. The exchange of information, through vertical messages, between detector and decoder can be iterated for $N_t$ times. 

With reference to the generic $n_{iter,T}$ turbo iteration, the proposed algorithm structure is outlined in Algorithm \ref{algorithm}.

\begin{algorithm}
	\caption{Proposed EP algorithm} 
        \label{algorithm}
	\begin{algorithmic}[1]
            \State \textbf{Initialization:}
            \State Initialize the messages $p^{(f)}_{0|-1}(\boldsymbol{h}_0)$ and $p^{(b)}_{K-1|K}(\boldsymbol{h}_{K-1})$ by setting $\boldsymbol{\Sigma}_{f,0}$, $\boldsymbol{\Sigma}_{b,K-1}$ to $\mathbb{I}_N$ and $\boldsymbol{\eta}_{0|-1}$, $\boldsymbol{\eta}_{K-1|K}$ to $\boldsymbol{0}$ according to the moments of the generated fading samples
		\For {$k = 0:K-2$}\label{FW_s}
                \State Compute the \textit{temporary prior} distribution, i.e., update its parameters $\boldsymbol{\eta}_{p,k}$ and $\boldsymbol{\Sigma}_{p,k}$ via~(\ref{eq:prior precision})-(\ref{eq:prior mean}). The backward message $p^{(b)}_{k|k+1}(\h)$ in~(\ref{eq:prior_distribution}) is set to the one computed at the ($n_{iter,T}-1$)-th turbo iteration. If $n_{iter,T}=1$, it is set to $1$ for any value of $k$\label{Alg:prior}
                \State Compute the parameters of the \textit{exact marginal} distribution (i.e., for each $m$, compute the terms in  (\ref{eq:P_m}), (\ref{eq:w_m})-(\ref{eq:J}))\label{Alg:MIX}
				\State Perform the \textit{moment matching}, using~(\ref{eq:eta_EP})-(\ref{eq:sigma_EP})\label{MM}
				\State Compute $\Tilde{p}_d(\h)$ via~(\ref{eq:Sigma_d})-(\ref{eq:eta_d}) and perform either \textit{damping} or \textit{boosting} on $\boldsymbol{\Sigma}_{d}$ depending on $k$
                \If{$(\boldsymbol{\Sigma}_{d}(1,1)<0)\And(n_{iter,T}>1)$}
                \State Perform the \textit{observation} message projection via~(\ref{eq:pd_proj_mean})-(\ref{eq:pd_proj_var})
                \EndIf\label{Proj}
                \State Compute the forward messages $p^{(f)}_{k|k}(\h)$  and $p^{(f)}_{k+1|k}(\boldsymbol{h}_{k+1})$ through the SPA rules~\cite{KsFrLo01} (refer to Fig.~\ref{fig:FG})\label{Alg: fw}
	   \EndFor\label{FW_f}
          \For {$k = K-1:-1:1$}\label{BW_s}
                \State Compute $p^{(p)}(\h)$ via~(\ref{eq:prior_distribution})-(\ref{eq:prior mean}) where $p^{(f)}_{k|k-1}(\h)$ is set to the one computed at the ($n_{iter,T}-1$)-th turbo iteration. If $n_{iter,T}=1$, it is set to $1$ for all $k$s
                \State Repeat the steps \ref{Alg:MIX}-\ref{Proj}
                \State Compute the backward messages $p^{(b)}_{k|k}(\h)$ and $p^{(b)}_{k-1|k}(\boldsymbol{h}_{k-1})$ through the SPA rules~\cite{KsFrLo01}\label{Alg: bw}
          \EndFor\label{BW_f}
          \State Multiply the forward message obtained through \ref{FW_s}-\ref{FW_f} with the backward one computed in \ref{BW_s}-\ref{BW_f} for each $k$\label{compl} 
          \State Compute for each $k$ and each $c_k\in \mathcal{X}$ the probabilities $P_u(c_k)$ to be sent to the decoder as
          \mbox{$P_u(c_k)\propto g_\mathbb{C}\left(r_k, c_k\boldsymbol{a}^{H}\boldsymbol{\eta}_{u,k}, N_0 + |c_k|^2\boldsymbol{a}^{H}\boldsymbol{C}_{u,k}\boldsymbol{a}\right)$}, where $\boldsymbol{\eta}_{u,k}$ and $\boldsymbol{C}_{u,k}$ derive from \ref{compl}
	\end{algorithmic} 
\end{algorithm}

Notice that the forward and backward recursions lead to two different approximations for ${p}_d(h_k)$. 
The main advantage of the adopted scheduling is that a satisfying performance is already achieved after a single forward-backward sweep inside the detector.

\section{Simulation Results}\label{sec:results}

\begin{figure}
    \centering
    \includegraphics[width=1\columnwidth]{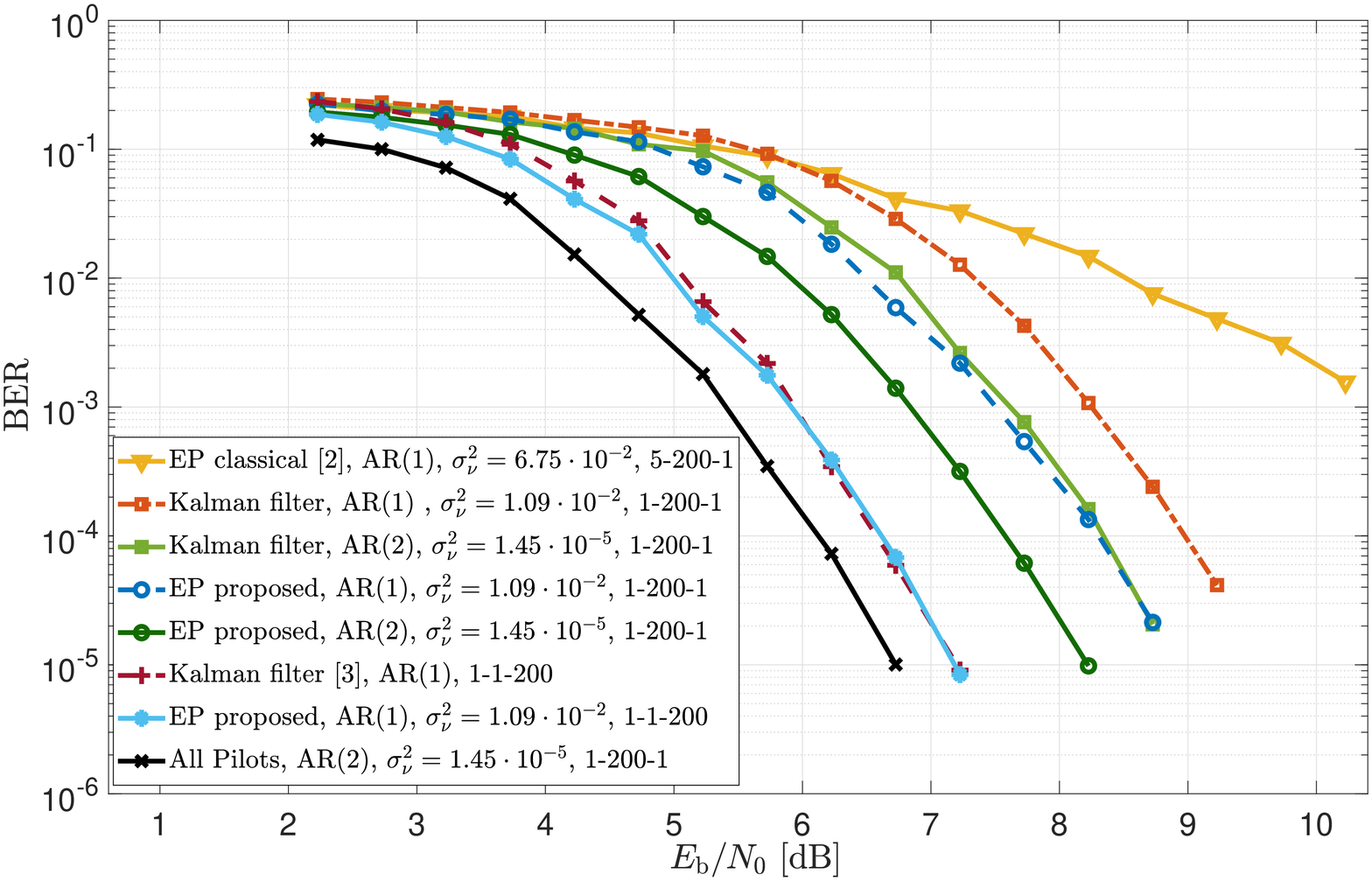}
    \caption{Performance for the $1/20$ pilots distribution with $f_DT=10^{-2}$.} 
    \label{fig:distributed_pilots}
\end{figure}

\begin{figure}
    \centering
    \includegraphics[width=1\columnwidth]{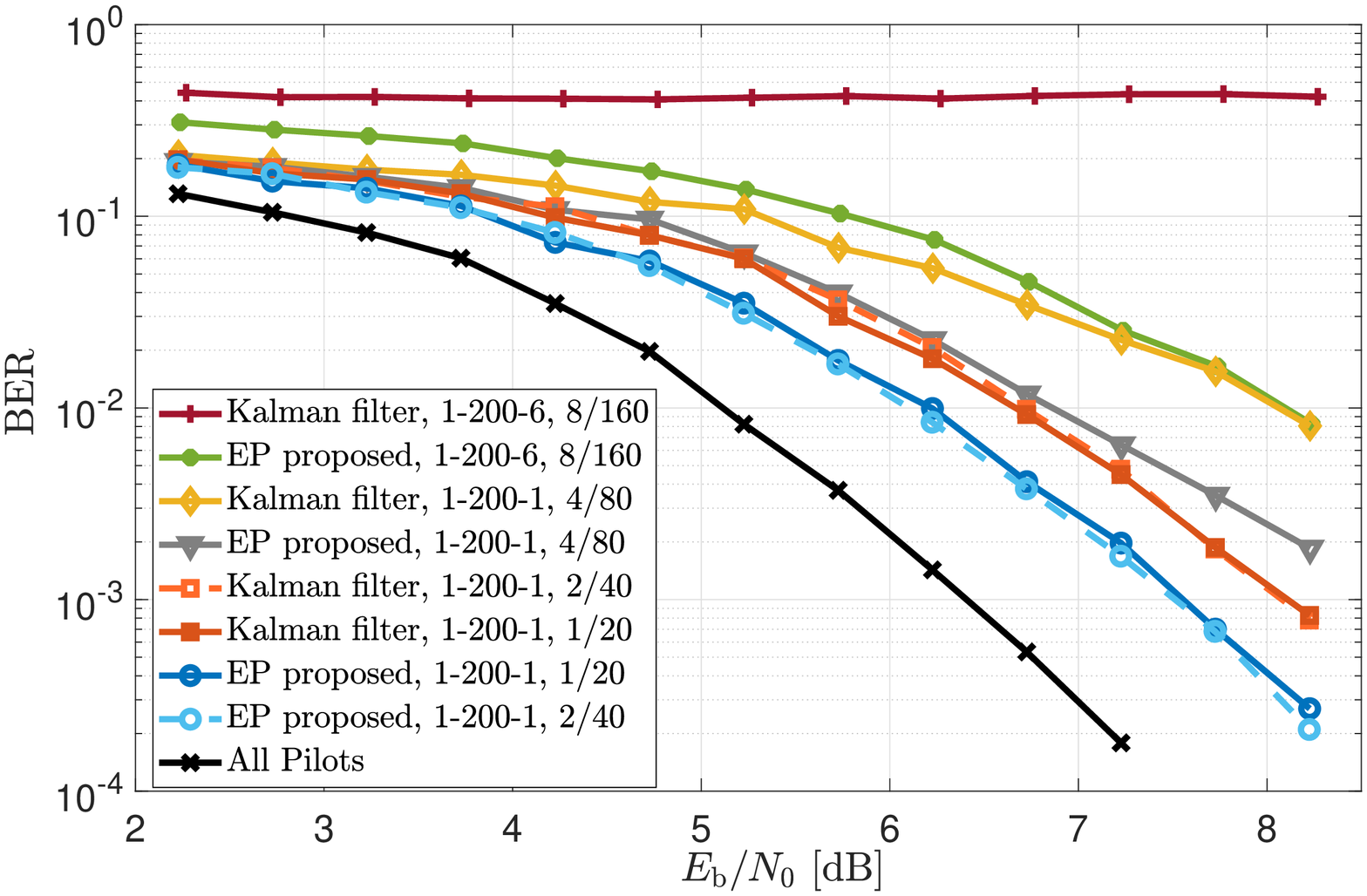}
    \caption{Performance for different pilots distribution with $f_DT=5\cdot10^{-3}$ and \mbox{$\sigma^2_\nu = 8\cdot 10^{-3}$}.}
    \label{fig:concentrated_pilots}
\end{figure}

The performance of the proposed algorithm was analyzed in terms of BER versus $E_b/N_0$, $E_b$ being the received signal energy per information bit and $N_0$ the one-sided noise power spectral density. We considered a $(3,6)$-regular rate-$1/2$ LDPC code with length $4000$ and QPSK modulation ($M=4$). %Simulations results for the case of perfect channel state information (CSI) at the receiver are shown for comparison. 
In all simulations, in order to allow the algorithms convergence, pilot symbols were inserted in the transmitted codeword.
\footnote{We use the notation $\{$number of pilots in a block$\}/\{$blocks distance$\}$ to describe a pilot distribution.}
The presence of pilots involves a slight decrease of the effective information rate, resulting in an increase in the required SNR. This increase has been artificially introduced in the curve labeled \textquotedblleft{All Pilots}\textquotedblright{\,}, inserted as ideal bound, for the sake of comparison. Hence, the gap between the \textquotedblleft{All Pilots}\textquotedblright{\,} and the other curves is uniquely due to the need for fading estimation/compensation, and not to the decrease due to pilot symbols. 
The curves labeled \textquotedblleft{Kalman filter}\textquotedblright{\,} in Figs.~\ref{fig:distributed_pilots} and~\ref{fig:concentrated_pilots} refer to the algorithm proposed in~\cite{BaPiCo09}.   

%FIGURE 2
The algorithms were tested for both the AR($1$) and AR($2$) fading models. The values of the parameter $\boldsymbol{\rho}$ were set according to the YW equations, whereas the increment variance $\sigma^2_\nu$ was set to the one minimizing the BER, found through computer simulations. 

\begin{figure}
    \centering
    \hspace*{-0.35cm}
    \includegraphics[width=1\columnwidth]{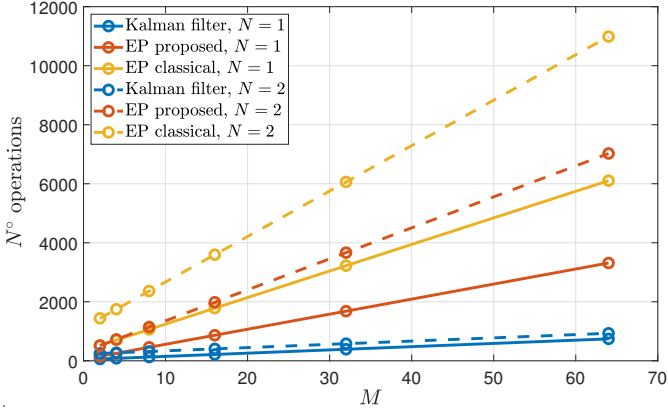}
    \caption{Detection algorithms complexity comparison for $N_d = 1$.}
    \label{fig:complexity}
\end{figure}

In Fig.~\ref{fig:distributed_pilots} we compared the performance of the proposed algorithm with both the Kalman filter and the \textquotedblleft{batch-EP}\textquotedblright{} of \cite{QiMi07}, labeled with \textquotedblleft{classical EP}\textquotedblright{}, when the normalized Doppler bandwidth of the fading process $f_DT$ was set to  $10^{-2}$ and one pilot symbol was inserted in every block of $20$ transmitted symbols. 
Firstly, we analyzed the practical case of separate detection and decoding.\footnote{In order to indicate the number of inner detector, inner decoder and turbo iterations, the notation $N_d$-$N_c$-$N_t$ is used.}
Figure~\ref{fig:distributed_pilots} shows that the proposed EP algorithm can achieve a performance gain of approximately $0.9\,$dB with respect to the Kalman filter for both the AR($1$) and the AR($2$) models.
Moreover, a significant improvement is observed comparing the proposed technique with the classical EP of~\cite{QiMi07}. Regarding the latter algorithm, the authors proposed an EP smoothing approach where an iterative refinement of the observation approximations is foreseen. 
Consistently with that used in~\cite{QiMi07}, a maximum number ($N_d$) of $5$ inner detector iterations was allowed, and it resulted to be sufficient for achieving convergence. For this algorithm, the curve shown in Fig.~\ref{fig:distributed_pilots} is obtained using the AR($1$) model since the AR($2$) model did not yield further improvement. The complexity of the considered algorithms was evaluated with reference to one inner detector iteration for a single time epoch $k$ and it is reported in Fig.~\ref{fig:complexity}. They were compared considering the total number of required operations, including two-terms multiplications and additions. The distinction among the various kind of operations is, instead, shown in Table~\ref{Tab:complexity} for $M=4$ where the number of Lookup Table (LUT) accesses for the computation of nonlinear terms is also specified. It can be noticed that the additional operations involved by the EP framework lead inevitably to a complexity increase which, relying on our implementation, is still limited and reasonably comparable to the one of the Kalman filter. On the other hand, a considerable complexity increase is registered for the classical strategy~\cite{QiMi07}.

\begin{table*}
\begin{center}
\begin{tabular}{|| c || c | c || c | c  || c ||}
    \hline 
    & \multicolumn{2}{c||}{\textbf{Two-terms additions}} &  \multicolumn{2}{c||}{\textbf{Two-terms multiplications}} & \textbf{LUT accesses} \\ [0.5ex]  
    \cline{2-6}
    & \hspace{0.25cm}$\boldsymbol{N=1}$\hspace{0.25cm} & $\boldsymbol{N=2}$ & \hspace{0.5cm}${\boldsymbol{N=1}}$\hspace{0.5cm} & ${\boldsymbol{N=2}}$ & $-$\\
    \hline\hline
    Kalman & ${\boldsymbol{24}}$ & ${\boldsymbol{97}}$ & ${\boldsymbol{53}}$ & ${\boldsymbol{177}}$ & $\boldsymbol{6}$   \\ 
    \hline
    EP proposed & ${\boldsymbol{86}}$ & ${\boldsymbol{261}}$  & ${\boldsymbol{148}}$ & ${\boldsymbol{452}}$ & $\boldsymbol{12}$    \\ 
    \hline
    EP classical & ${\boldsymbol{217}}$ & ${\boldsymbol{602}}$ & ${\boldsymbol{478}}$ & ${\boldsymbol{1138}}$ & $\boldsymbol{8}$    \\ 
    \hline
    \end{tabular}
    \vspace{0.1cm}
    \caption{\label{Tab:complexity} Complexity comparison for $M=4$ and $N_t = 1$.}
\end{center}
\end{table*}

In Fig.~\ref{fig:distributed_pilots}, the BER curves related to the joint iterative detection and decoding, where a maximum of $200$ turbo iterations is allowed, are also reported with reference to the AR($1$) fading model. It is possible to notice that the curves related to the proposed EP algorithm and the Kalman filter (optimized according to~\cite{BaPiCo09}) are approximately overlapped. In fact, when a sufficient number of turbo iterations is performed, the decoder provides symbol probabilities approaching the ones passed in correspondence with a known symbol. Therefore, the observation message is composed by a single Gaussian and the operations performed by the two detectors coincide. 

In Fig.~\ref{fig:concentrated_pilots}, different pilot distributions were considered in order to show the higher robustness of EP with respect to the Kalman filter against pilot symbols arrangements in sequences separated by longer blocks of code symbols. 
In particular, the analyzed configuration are $1/20$, $2/40$, $4/80$, $8/160$, so that the effective information rate is the same.
The normalized Doppler bandwidth was set to $5\cdot 10^{-3}$. The AR($1$) fading model was adopted for the sake of simplicity, however, similar conclusions can be drawn for AR($2$). 
It is possible to notice that both the proposed algorithm and the Kalman filter are practically insensitive to a change in the pilots blocks distance from $20$ to $40$ symbols. Differently, increasing the gap to $80$ symbols 
causes a loss of about $0.8\,$dB for the EP-based algorithm and about $1.5\,$dB for the Kalman filter approach.
Finally, considering a pilots distance of $160$ code symbols, the EP detector is still able to provide reasonable fading estimates and, with the aid of $6$ turbo iterations, it loses approximately $2\,$dB from the case where the $1/20$ pilots distribution is adopted. On the contrary, in the same conditions, the Kalman filter does not work because of its lower robustness against longer pilots spacing. 

\begin{figure}
    \centering
    \hspace*{-0.2cm}
    \includegraphics[width=1.05\columnwidth]{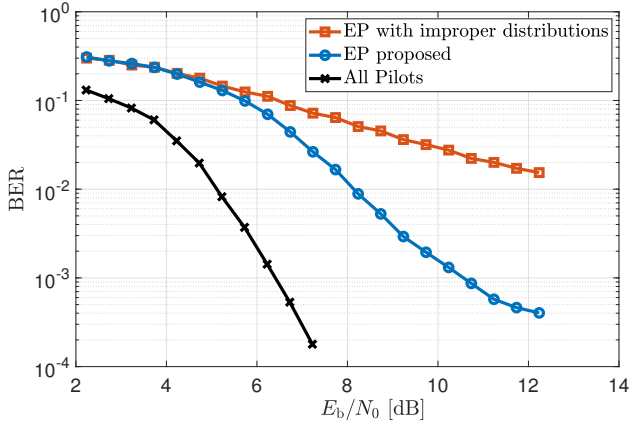}
    \caption{Performance comparison of the EP algorithm adopting the strategy proposed in Section~\ref{subsec:neg_variances} with the procedure based on the propagation of improper distributions. The analyzed scenario is based on the concentrated pilots configuration, in particular the $8/160$ one, and the normalized Doppler bandwidth $f_d T = 5\cdot 10^{-3}$. The selected $\sigma^2_\nu$ optimizing the performance is set to $8\cdot  10^{-3}$. The simulated iterations are $1$-$200$-$7$.}
    \label{fig:improper_distributions}
\end{figure}

The proposed solution in this last situation adopts the observation message projection strategy presented in Section~\ref{subsec:neg_variances}. In order to show the advantages of this approach, we compared the algorithm performance with and without the handling of improper distributions in Fig.~\ref{fig:improper_distributions}. It can be noted that a considerable gain can be achieved by blocking the improper distributions propagation. This is due to the fact that the adopted pilots configuration makes the algorithm operating conditions harder. Therefore, the estimate obtained by the \textit{prior} is not often in accordance with any of the observation message modes, leading to a rather wide approximating marginal distribution. Negative variances are thus typically generated after the observation message update, and allowing their propagation and combination with those from previous iterations leads to poor performance. 
On the other hand, by considering more favourable conditions, the performance gap becomes less relevant.

% ************************************************

\section{Conclusions}\label{sec:conclusions}

In this paper, we proposed a new solution for signal detection over channels with flat-fading through EP. The presented method allows to solve the numerical instabilities related to the traditional procedure provided in the literature. We showed through numerical results the advantages in terms of BER that can be achieved in the separate detection and decoding case with respect to the classical Kalman filter approach and we demonstrated the higher robustness of this algorithm against more concentrated pilots distributions.

The advantages of this framework offer a variety of possibilities for new research directions, such as its application to the multiple-input multiple-output channel model and the study of more severe fading conditions, for instance frequency selectivity. Moreover, we could extend our analysis to other kinds of divergence measures for the involved approximations.

\bibliographystyle{IEEEtran}

% Generated by IEEEtran.bst, version: 1.14 (2015/08/26)

\end{document}